# Modeling Aftershocks as a Stretched Exponential Relaxation

A. Mignan

Institute of Geophysics, Swiss Federal Institute of Technology Zurich, Switzerland

arnaud.mignan@sed.ethz.ch

**Abstract**: The decay rate of aftershocks has been modeled as a power law since the pioneering work of Omori in the late nineteenth century. Considered the second most fundamental empirical law after the Gutenberg-Richter relationship, the power law paradigm has rarely been challenged by the seismological community. By taking a view of aftershock research not biased by prior conceptions of Omori power law decay and by applying statistical methods recommended in applied mathematics, I show that all aftershock sequences tested in three regional earthquake catalogs (Southern and Northern California, Taiwan) and with three declustering techniques (nearest-neighbor, second-order moment, window methods) follow a stretched exponential instead of a power law. These results infer that aftershocks are due to a simpler relaxation process than originally thought, in accordance with most other relaxation processes observed in Nature.

## 1. Introduction

Since the first description of the aftershock decay rate $n(t) \propto (t + t_{min})^{-1}$ as a power law by Omori in 1894 [Omori, 1894], the only notable modification made to the model has been the establishment of the generalized form $n(t) \propto (t + t_{min})^{-\alpha}$ in the mid-twentieth century by Utsu [Utsu, 1961] (note that similar versions had already been proposed by Hirano [1924] and Jeffreys [1938] in more obscure journals). The so-called Modified Omori law (MOL, or Omori-Utsu law) now forms the basis for the widely used Epidemic-Type Aftershock Sequence (ETAS) model [Ogata, 1983; 1988; Helmstetter and Sornette, 2002]. Although alternative models have been suggested to describe the aftershock decay rate [e.g., Souriau et al., 1982; Kisslinger, 1993; Gross and Kisslinger, 1994; Narteau et al., 2002; Lolli and Gasperini, 2006], the few available model comparisons have shown mixed results as to their performance relative to the MOL [Kisslinger, 1993; Gross and Kisslinger, 1994; Narteau et al., 2002; Lolli and Gasperini, 2006].

Physical relaxation processes are most often described by a pure exponential decay of the form $n(t) \propto \exp(-\lambda t)$ [Leike, 2002] or by a stretched exponential of the form $t^{\beta-1}exp(-\lambda t^{\beta})$ where $0 < \beta < 1$ [Phillips, 1996; Laherrère and Sornette, 1998] (Note that $\beta = 1$ yields the pure exponential function). The suggestion that aftershock decay would follow a power law instead of an expression from the exponential family inspired early criticisms of the Omori law. As indicated by Burridge and Knopoff [1967], "*Richter (1958) has noted that an exponential formula would be more plausible on physical grounds*". Utsu et al. [1995] added in his review that "*the power law implies the long-lived nature of activity in contrast to the exponential function appearing in most decay laws in physics*". The apparent power law behavior has been explained by a combination of exponential decays, assuming that $\lambda$ is itself exponentially distributed [Lomnitz, 1974], as well as by special cases of the rate-and-state [Dieterich, 1994] and damage rheology [Ben-Zion and Lyakhovsky, 2006] theories.

The aim of the present study is to demonstrate that aftershock decay is better described by a stretched exponential than by a power law when one takes a view of aftershock research not biased by prior conceptions of Omori power law decay. First, I make a meta-analysis of the aftershock decay models published from 1894 until present. Such an approach has already been shown to offer valuable information on the scientific process leading to the description of earthquake populations (e.g., preshock and foreshock cases) [Mignan, 2011; 2014]. Second, I systematically compare the power law, exponential and stretched exponential in their simplest form to describe the aftershock decay rate observed in three regional earthquake catalogues: Southern California, Northern California and Taiwan. Aftershock sequences are defined from three different declustering techniques for sensitivity analysis [Gardner and Knopoff, 1974; Reasenberg, 1985; Zaliapin et al., 2008]. I use the mathematical formulations and statistical methods recommended in recent years in applied mathematics [Clauset et al., 2009] in which the power law differs from the Omori law formulation.

**2. Meta-analysis of aftershock decay rate models**

Twenty aftershock decay rate models published between 1894 and 2006 are uncovered from a review of the seismological literature (Table 1). Figure 1 shows the relationships between these decay rate models $n(t)$ as a function of publication year and number of parameters. Models that are linked indicate that the one with fewer parameters is nested in the other one. A combination of all linked models then forms a model family. The popularity of a formula is estimated by the number of times it has been cited (represented by the circle size in Fig. 1).

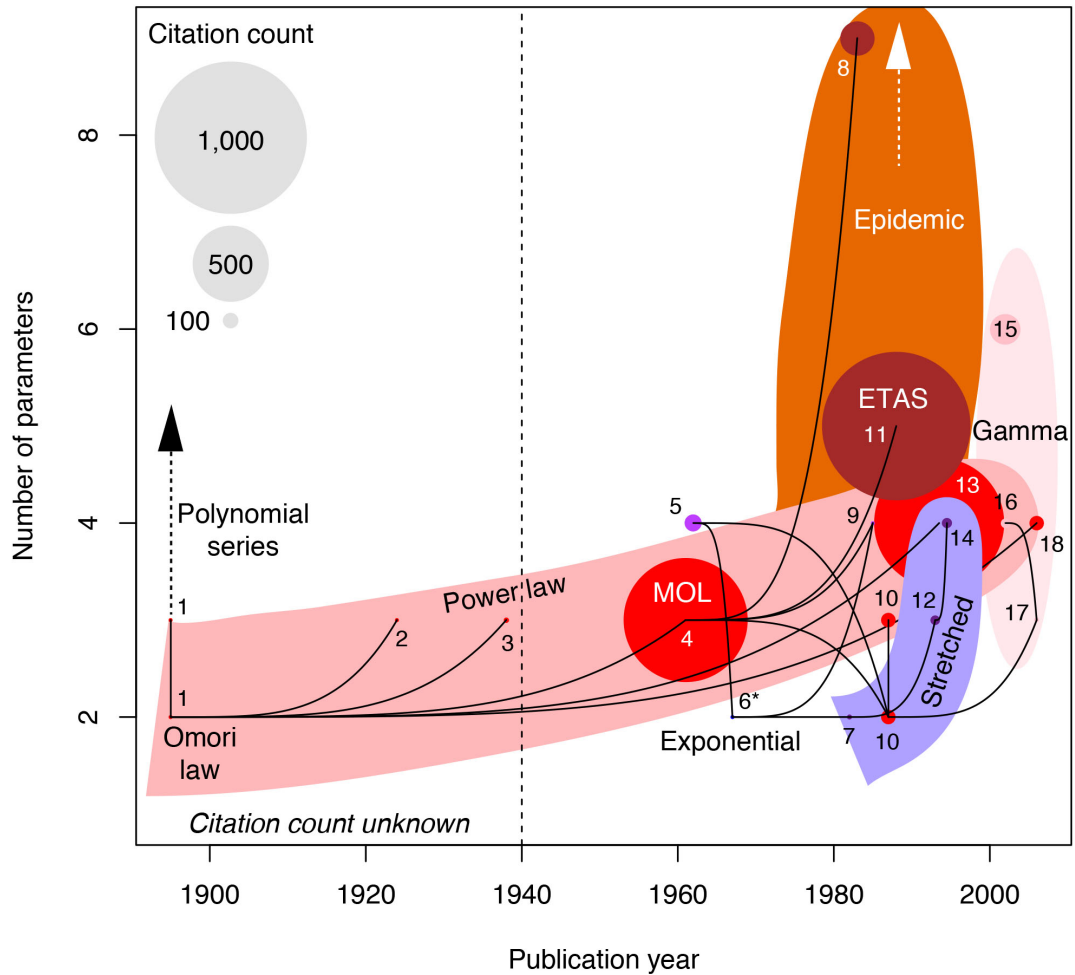

**Fig. 1**. Meta-analysis of aftershock decay rate formulas suggested in the literature. Each numbered solid circle represents one publication (see Table 1). Nested formulas are connected by curves. Different model families or sub-families (power law, epidemic extension, stretched exponential and other power-law/exponential hybrids, gamma-related) are represented by different colors. Citation count obtained from *GoogleScholar* (as of July 2015) except for Burridge and Knopoff [1967], whose reference to an exponential behavior is only anecdotal and the citation count not related to the formula.

The power law family is the most highly cited with the Utsu [1962] (MOL) formulation cited more than 800 times and its epidemic extension ETAS [Ogata, 1988] cited almost 1,000 times. The stretched exponential family, which emerged in the 1980's, remains marginal with about 50 citations per suggested model [Souriau et al., 1982; Kisslinger, 1993; Gross and Kisslinger, 1994]. Models based on the Gamma function started to be suggested only in the 2000's [Helmstetter and Sornette, 2002; Narteau et al., 2002; Lolli and Gasperini, 2006]. Some other models have been proposed based on a mixture of a power law and an exponential [Mogi, 1962; Otsuka, 1985]. The pure exponential was however tested only once and on one aftershock sequence only [Burridge and Knopoff, 1967]. Although the rate-and-state and damage rheology theories [Dieterich, 1994; Ben-Zion and Lyakhovsky, 2006] have also proposed aftershock decay models defined in exponential terms, they are rarely considered for aftershock fitting (and thus only their power law variants shown in Fig. 1 and Table 1). Lolli et al. [2009] did not consider the Dieterich formulation in their aftershock model comparison indicating it was "*not particularly well suited to reproduce real sequences*". In compliance with the Omori law, Yang and Ben-Zion [2009] only considered the aftershock power law version derived from the damage rheology approach. Note that $n(t,m)$ models, which additionally include the role of the magnitude $m$, are not considered in the present study but often also consider a temporal term based on the MOL [e.g., Shcherbakov et al., 2004].

Although Figure 1 indicates a general agreement for using aftershock models of the MOL family (see also discussions in Utsu et al. [1995] and Vere-Jones [2000]), attention must be pointed to some of its known limitations. First, the addition of a starting time $t_{min}$ in $n(t)$ is unphysical. First suggested to improve data fitting [Omori, 1894], this parameter avoids the singularity at $t = 0$. The interval $[0, t_{min}]$ has been shown to correspond to a period during which the aftershock sequence is incomplete [e.g, Lolli and Gasperini, 2006]. This is problematic since $t_{min}$ can be correlated with the power exponent α [Lolli and Gasperini, 2006] and since power law behavior should only be tested on the interval $[t_{min}, +\infty)$ [e.g., Clauset et al., 2009; Newman, 2005]. Second, the case α ≤ 1 requires aftershock sequences to be bounded at $t_{max}$ to avoid an infinite area under the tail. The *ad hoc* parameter $t_{max}$ has also no physical meaning. Although some studies have obtained α ≤ 1 [e.g., Kisslinger, 1993; Utsu et al., 1995; Lolli and Gasperini, 2006], this case is virtually never observed in Nature [Newman, 2005]. For the case α > 1 and for the exponential family, the rate of aftershocks naturally tends to zero toward infinity.

Another type of limitation, which applies to all model families, relates to the ambiguous definition of aftershocks. Earthquakes are identified as aftershocks if they follow specific clustering rules, which depend on the declustering technique that is chosen [van Stiphout et al, 2012]. *A priori*, different techniques could lead to different temporal forms of the aftershock decay. So far, all aftershock model comparisons have been based on one aftershock definition. The window method of Gardner and Knopoff [1974] is used in Narteau et al. [2002] and Lolli and Gasperini [2006] while the second-order moment method of Reasenberg [1985] is used in Kisslinger [1993] and Gross and Kisslinger [1994].

**3. Aftershock temporal decay fitting method**

Aftershock sequences are identified using three different declustering methods, the nearest-neighbor method of Zaliapin et al. [2008], the second-order moment method of Reasenberg [1985] and the window method of Gardner and Knopoff [1974]. They are all used with their standard parameterization. I use the parameter sets {$d$ = 1.6, $b$ = 1.0, $p$ = 0.5} from Zaliapin et al. [2008] as given in their MATLAB code and {$\tau_{min}$ = 1 day, $\tau_{max}$ = 10 days, $p_1$ =

0.95, $x_k$ = 0.5, $x_{meff}$ = 1.5, $r_{fact}$ = 10} from Reasenberg [1985] as given in van Stiphout et al. [2012]. For the window method, I use the spatial window $10^{0.1238m+0.983}$ in km and the temporal windows $10^{0.032m+2.7389}$ in days if $m \geq 6.5$ and $10^{0.5409m-0.547}$ otherwise [van Stiphout et al, 2012; Gardner and Knopoff, 1974]. van Stiphout et al. [2012] showed that inter-technique uncertainties are greater than intra-technique uncertainties. For this reason, I do not investigate the impact of different technique parameterizations. The three declustering methods are tested on the three following relocated earthquake catalogues: Southern California (1981-2011), Northern California (1984-2011) and Taiwan (1991-2005) [Hauksson et al., 2012; Waldhauser and Schaff, 2008; Wu et al., 2008]. Only aftershock sequences composed of more than 100 events with magnitude $m \geq m_{min}$ are considered, $m_{min} \geq 2.0$ being the completeness magnitude of each sequence.

The magnitude and temporal lower bounds $m_{min}$ and $t_{min}$ are estimated successively, for each aftershock sequence, using an approach that combines maximum-likelihood fitting with goodness-of-fit tests based on the Kolmogorov-Smirnov (KS) statistic [Clauset et al., 2009]. First the predicted rate $n_{pred}(m)$ is computed over the magnitude range [$m_i$, +∞), for each minimum magnitude $m_i$ defined in the interval [$m(\max n_{obs})$, $\max m$], with the magnitude bin $\Delta m$ = 0.1 and $\log_{10}(n_{pred}(m))$ = $-b_{MLE}(m-\Delta m/2)$ the Gutenberg-Richter relationship for an aftershock sequence over its entire temporal domain. The lower bound $m(\max n_{obs})$ corresponds to the lowest possible completeness magnitude [Mignan, 2012]. The parameter $b_{MLE}$ = $\log_{10}(e)/(\bar{m}-\Delta m/2)$ is the maximum likelihood estimate of the slope of the Gutenberg-Richter law [Aki, 1965]. Then the lower magnitude bound $m_{min}$ is defined as the estimate $m_i$ for which the distance $D$ = $\max_{m \geq mi}$ | CDF($n_{obs}(m)$) - CDF($n_{pred}(m)$) | is minimized (i.e., KS test with CDF the cumulative distribution function). The resulting $m_{min}$ indicates the magnitude bin above which the Gutenber-Richter law is verified (see list of median $m_{min}$ values per catalogue and declustering method in Table 2). This estimate represents the long-term completeness of the sequence while the lower temporal bound $t_{min}$ represents the time after which the seismic network is not anymore saturated by the effects of the mainshock. $t_{min}$ is estimated following the same approach as for $m_{min}$ but with $n_{pred}(t, m \geq m_{min})$ the power law, exponential or stretched exponential function defined in their simplest form: $n(t) \propto t^{-\alpha}$, $n(t) \propto \exp(-\lambda t)$ and $n(t) \propto t^{\beta-1} exp(-\lambda t^\beta)$ [Clauset et al., 2009] (note that the simple power law formulation was also used in Yamashita and Knopoff [1987]). It means that for each sequence, three $t_{min}$ estimates are obtained, each to optimize one model. To *a priori* not favor any model, the three models are compared for each optimized $t_{min}$ (i.e. on the same datasets) (median $t_{min}$ values per catalogue, declustering method and optimized model are given in Table 2). In average, $t_{min}$ is lower when optimized for the stretched exponential instead of the power law, which indicates that the stretched exponential applies on a greater range.

The best fit to each aftershock sequence is determined using the Akaike Information Criterion AIC = $2k-2LL$ [Akaike, 1974] where $k$ is the number of parameters and $LL$ the log-likelihood of the aftershock decay rate model for $m \geq m_{min}$ and $t \geq t_{min}$. For both the power law and the exponential, $k = 1$ (parameter $\alpha$ or $\lambda$, respectively). For the stretched exponential, $k = 2$ (parameters $\beta$ and $\lambda$). The log-likelihood is computed from the PDF of the power law, exponential and stretched exponential functions, as defined in Clauset et al. [2009]. The best model per sequence and per model-optimized $t_{min}$ is the model with the lowest AIC.

## 4. Systematic aftershock model comparison

Due to the known limitations of the MOL and to the lack of systematic model comparison, I test the power law, exponential and stretched exponential on the interval $[t_{min}, +\infty)$ using the method described above. Figure 2 shows all the aftershock sequences considered (grey curves) and the median fits for the power law and stretched exponential (red and purple, respectively - the pure exponential corresponding to the stretched exponential with $\beta = 1$). Table 2 shows the percentage of sequences best fitted by a given model per region, per declustering method and per model-optimized $t_{min}$. In all cases (total of 245 sequences), the aftershock decay rate is best described by a stretched exponential or pure exponential rather than by a power law. If $t_{min}$ is optimized for the power law, at most 6% of the sequences are best explained by it. However all of these sequences are best fitted by the stretched exponential or pure exponential if $t_{min}$ is optimized for these respective models. The stretched exponential performs better than the exponential in all regions and declustering methods except only for the Taiwan region when aftershocks are defined with the window method. For aftershock sequences defined from the nearest-neighbor or second-order moment methods, the stretched exponential best explains the data between 85 and 100% of cases (and the exponential the remaining sequences). For sequences defined from the window method, the stretched exponential best explains the data between 34 and 74% of cases (and the exponential the remaining sequences).

Parameters fluctuate more across declustering methods than across regions, with median values of $\alpha = 1.34$-$1.41$, $\beta = 0.30$-$0.38$ and $\lambda = 0.50$-$0.57$ for the nearest-neighbor method, $\alpha = 1.46$-$1.53$, $\beta = 0.52$-$0.62$ and $\lambda = 0.47$-$1.00$ for the second-order moment method and $\alpha = 1.30$-$1.47$, $\beta = 0.82$-$1.00$ and $\lambda = 0.03$-$0.27$ for the window method. The parameter variability reflects the ambiguity in defining aftershocks. Nonetheless, the exponential family (stretched or pure) is preferred to the power law whichever definition is used.

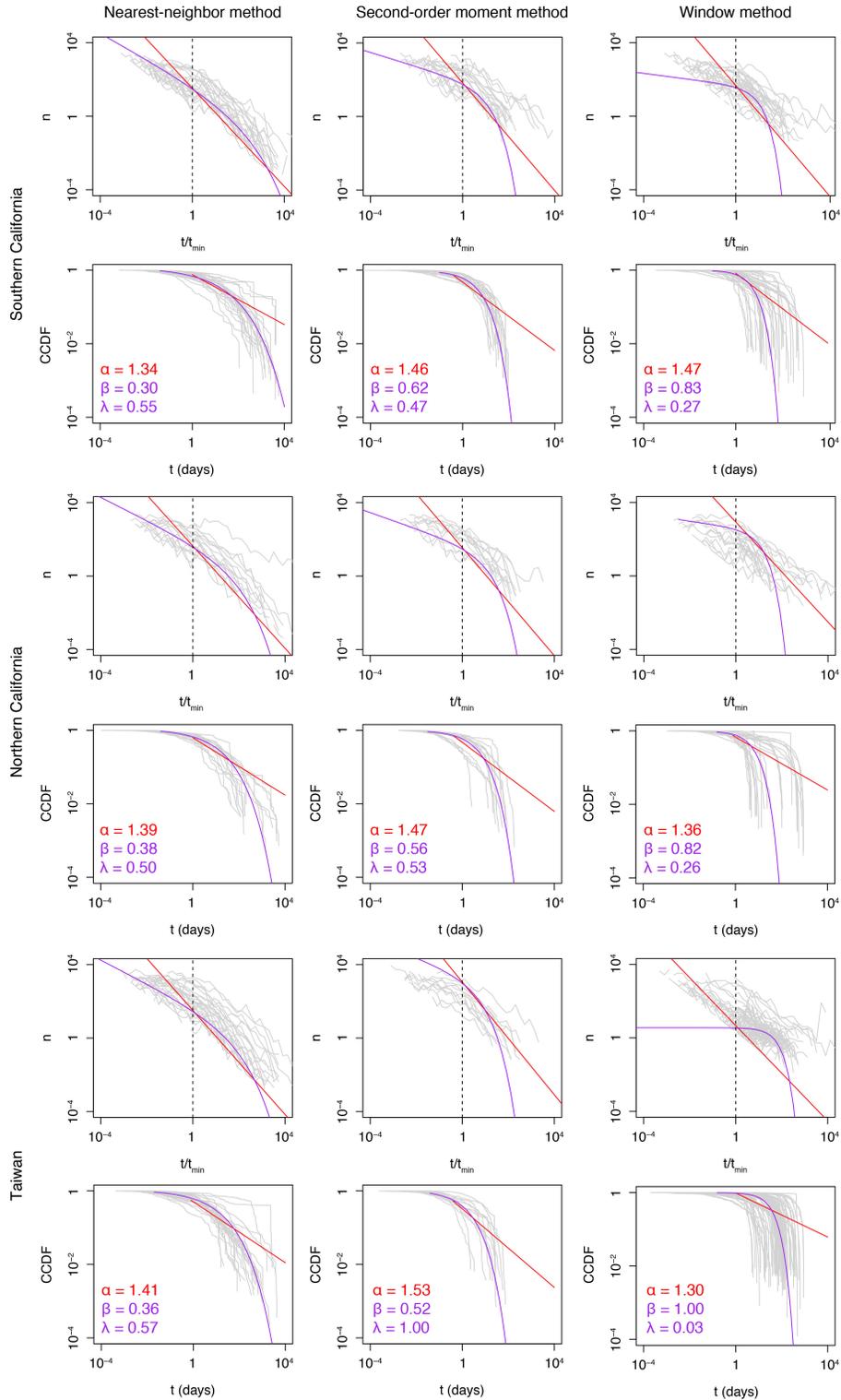

**Fig. 2.** Aftershock sequence stacks and median fits. Sequences (grey curves) are defined by three declustering methods in three regional catalogues. The aftershock decay rate $n(t)$ log-log plot commonly used in statistical seismology may suggest a power law behavior (apparent straight line above $t/t_{min} = 1$ with $t_{min}$ optimized for the power law model). The CCDF log-log plot shows instead a clear curvature. Power law and stretched exponential fits (red and purple, respectively) are shown for median parameter estimates only. A pure exponential is obtained when $\beta = 1$. Median values are given for $\alpha$, $\beta$ and $\lambda$ per region and per declustering method.

There are at least reasons that past studies failed to observe the stretched exponential decay rate of aftershocks. First, the MOL is generally not challenged due to the general agreement that this formulation is correct. This idea is supported by the apparent linear behavior of the rate $n(t \geq t_{min})$ in a log-log plot (Fig. 2). Although commonly used in aftershock studies [e.g., Utsu et al., 1995; Helmstetter and Sornette, 2002], it has been recommended to plot the complementary cumulative density function (CCDF) instead to identify possible deviations from a power law behavior [e.g., Clauset et al., 2009]. Figure 2 clearly shows that the apparent straight line observed in the $n(t)$ log-log plot disappears when replaced by the CCDF. This confirms that aftershocks do not follow a power law. Second, the few existing model comparisons [Kisslinger, 1993; Gross and Kisslinger, 1994; Narteau et al., 2002; Lolli and Gasperini, 2006] may have obtained mixed results because of the combined maximum likelihood estimation of $t_{min}$ with other parameters and/or of the use of relatively complex stretched exponential formulations [Souriau et al., 1982; Kisslinger, 1993; Gross and Kisslinger, 1994] and/or of the use of a relatively small number of aftershock sequences. Third, the present study is the first to use the methods proposed in recent years in applied mathematics to properly identify when empirical data follow or deviate from a power law [Newman, 2005; Clauset et al, 2009] and to do such tests on three regional catalogues with three different declustering methods yielding a total of 245 tested aftershock sequences.

5. Conclusions

I have shown that all aftershock sequences defined by three declustering techniques in Southern California, Northern California and Taiwan, are best described by a stretched exponential than by a power law. These results infer that the relaxation process that drives aftershocks is simpler than has been believed for decades. While a power law would require aftershock sequences to be the sum of exponential decays with their rate parameter λ exponentially distributed [Lomnitz, 1974], the present study shows that any aftershock sequence is described by one unique rate parameter λ. Most importantly, the stretched exponential function already describes most relaxation data observed in Nature [Laherrère and Sornette, 1998] and is often considered a universal property of relaxing systems [Phillips, 1996]. Compared to the pure exponential decay, the stretched exponential indicates that the decay rate is not constant but decreases with time as $t^{\beta-1}$ (Kohlrausch relaxation). This has been explained theoretically for electronic and molecular glasses as well as for Ising systems [Phillips, 1996; Huse et al., 1987]. It would suggest that the Earth's crust behaves similarly to any other homogeneously disordered solid. The values of the parameter β are here sensitive to the method from which aftershocks are defined (Fig. 2, Table 2). Therefore it is not yet possible to relate the aftershock crustal β-value to theoretical conditions [e.g., Phillips, 1996].

**References**

Akaike, H. (1974), A New Look at the Statistical Model Identification, *IEEE Trans. Automatic Control*, AC-19, 716-723

Aki, K. (1965), Maximum Likelihood Estimate of b in the Formula log N = a-bM and its Confidence Limits, *Bull. Earthquake Res. Instit.*, 43, 237-239

Ben-Zion, Y. and V. Lyakhovsky (2006), Analysis of aftershocks in a lithospheric model with seismogenic zone governed by damage rheology, *Geophys. J. Int.*, 165, 197-210, doi: 10.1111/j.1365-246X.2006.02878.x

Burridge, B. and L. Knopoff (1967), Model and theoretical seismicity, *Bull. Seismol. Soc. Am.*, 57, 341-371

Clauset, A., C. R. Shalizi, and M. E. J. Newman (2009), Power-Law Distributions in Empirical Data, *SAM Review*, 51, 661-703, doi: 10.1137/070710111

Dieterich, J. A. (1994), Constitutive law for rate of earthquake production and its application to earthquake clustering, *J. Geophys. Res.*, 99, 2,601-2,618

Gardner, J. K., and L. Knopoff (1974), Is the sequence of earthquakes in Southern California, with aftershocks removed, Poissonian?, *Bull. Seismol. Soc. Am.*, 64, 1,363-1,367

Gross, S. J., and C. Kisslinger (1994), Tests of Models of Aftershock Rate Decay, *Bull. Seismol. Soc. Am.*, 84, 1571-1579

Hauksson, E., W. Yang, and P. M. Shearer (2012), Waveform Relocated Earthquake Catalog for Southern California (1981 to June 2011), *Bull. Seismol. Soc. Am.*, 102, 2,239-2,244, doi: 10.1785/0120120010

Helmstetter, A., and D. Sornette (2002), Subcritical and supercritical regimes in epidemic models of earthquake aftershocks, *J. Geophys. Res.*, 107, 2237, doi: 10.1029/2001JB001580

Hirano, R. (1924), Investigation of aftershocks of the great Kanto earthquake at Kumagaya, *Kishoshushi Ser. 2* (in Japanese), 2, 77-83

Huse, D. A., and D. S. Fisher (1987), Dynamics of droplet fluctuations in pure and random Ising systems. *Phys. Rev. B*, 35, 6,841-6,848

Jeffreys, H. (1938), Aftershocks and periodicity in earthquakes, *Gerlands Beitr. Geophys.*, 56, 111-139

Kisslinger, C. (1993), The Stretched Exponential Function as an Alternative Model for Aftershock Decay Rate, *J. Geophys. Res.*, 98, 1,913-1,921

Laherrère, J., and D. Sornette (1998), Stretched exponential distributions in nature and economy: "fat tails" with characteristic scales, *Eur. Phys. J. B*, 2, 525-539

Leike, A. (2002), Demonstration of the exponential decay law using beer froth, *Eur. J. Phys.*, 23, 21-26

Lolli, B., and P. Gasperini (2006), Comparing different models of aftershock rate decay: The role of catalog incompleteness in the first times after main shock, *Tectonophysics,* 423, 43-59, doi: 10.1016/j.tecto.2006.03.025

Lolli, B, E. Boschi and P. Gasperini (2009), A comparative analysis of different models of aftershock rate decay by maximum likelihood estimation of simulated sequences, *J. Geophys. Res.*, 114, B01305, doi: 10.1029/2008JB005614

Lomnitz, C. (1974), Global Tectonics and Earthquake Risk, *Elsevier* London, 334 pp.

Mignan, A. (2011), Retrospective on the Accelerating Seismic Release (ASR) hypothesis: Controversy and new horizons, *Tectonophysics*, 505, 1-16, doi: 10.1016/j.tecto.2011.03.010

Mignan, A. (2012), Functional shape of the earthquake frequency-magnitude distribution and completeness magnitude, *J. Geophys. Res.*, 117, B08302, doi: 10.1029/2012JB009347

Mignan, A. (2014), The debate on the prognostic value of earthquake foreshocks: A meta-analysis, *Sci. Rep.,* 4, 4099, doi; 10.1038/srep04099

Mogi, K. (1962), Study of Elastic Shocks Caused by the Fracture of Heterogeneous Materials and its Relations to Earthquake Phenomena, *Bull. Earthquake Res. Instit.*, 40, 125-173

Narteau, C., P. Shebalin, and M. Holschneider (2002), Temporal limits of the power law aftershock decay rate, *J. Geophys. Res.*, 107, 2359, doi: 10.1029/2002JB001868

Newman, M. E. J. (2005), Power laws, Pareto distributions and Zipf's law, *Contemporary Physics*, 46, 323-351, doi: 10.1080/00107510500052444

Ogata, Y. (1983), Estimation of the parameters in the Modified Omori formula for aftershock frequencies by the maximum likelihood procedure, *J. Phys. Earth*, 31, 115-124

Ogata, Y. (1988), Statistical models for earthquake occurrences and residual analysis for point processes, *J. Am. Stat. Assoc.,* 83, 9-27

Omori, F. (1894), On after-shocks of earthquakes, *J. Coll. Sci. Imp. Univ. Tokyo*, 7, 111-200

Otsuka, M. (1985), Studies on aftershock sequences - Part 1. Physical interpretations of Omori's formula, *Sci Rep. Shimabara Earthq. Volcano Obs.* (in Japanese), 13, 11-20

Phillips, J. C. (1996), Stretched exponential relaxation in molecular and electronic glasses, *Rep. Prog. Phys.*, 59, 1,133-1,207

Reasenberg, P. (1985), Second-Order Moment of Central California Seismicity, 1969-1982, *J. Geophys. Res.*, 90, 5479-5495

Shcherbakov, R., D. L. Turcotte and J. B. Rundle (2004), A generalized Omori's law for earthquake aftershock decay, *Geophys. Res. Lett.*, 31, L11613, doi: 10.1029/2004GL019808

Souriau, M., A. Souriau, and J. Gagnepain (1982), Modeling and detecting interactions between earth tides and earthquakes with applications to an aftershock sequence in the Pyrenees, *Bull. Seismol. Soc. Am.*, 72, 165-180

Utsu, T. A. (1961), Statistical study on the occurrence of aftershocks, *Geophys. Mag.*, 30, 521-605

Utsu, T., Y. Ogata, and R. Matsu'ura (1995), The Centenary of the Omori Formula for a Decay Law of Aftershock Activity, *J. Phys. Earth*, 43, 1-33

van Stiphout, T., J. Zhuang and D. Marsan (2012), Seismicity declustering, *Community Online Resource for Statistical Seismicity Analysis*, doi: 10.5078/corssa-52382934, available at http://www.corssa.org

Vere-Jones, D. (2000), Seismology - A Statistical Vignette, *J. Am. Stat. Assoc.*, 95, 975-978, doi: 10.1080/01621459.2000.10474288

Waldhauser, F., and D. P. Schaff (2008), Large-scale relocation of two decades of Northern California seismicity using cross-correlation and double-difference methods, *J. Geophys. Res.*, 113, B08311

Wu, Y.-M., C.-H. Chang, L. Zhao, T.-L. Teng, and M. Nakamura (2008), A Comprehensive Relocation of Earthquakes in Taiwan from 1991 to 2005, *Bull. Seismol. Soc. Am.*, 98, 1,471-1,481, doi: 10.1785/0120070166

Yamashita, T., and L. Knopoff (1987), Models of aftershock occurrence, *Geophys. J. R. astr. Soc.*, 91, 13-26

Yang, W. and Y. Ben-Zion (2009), Observational analysis of correlations between aftershock productivities and regional conditions in the context of a damage rheology model, *Geophys. J. Int.*, 177, 481-490, doi: 10.1111/j.1365-246X.2009.04145.x

Zaliapin, I., A. Gabrielov, V. Keilis-Borok, and H. Wong (2008), Clustering Analysis of Seismicity and Aftershock Identification, *Phys. Rev. Lett.*, 101, 018501, doi: 10.1103/PhysRevLett.101.018501

**Table 1.** Aftershock decay rate formulas in literature.

| No. / Ref. | Formula $n(t) =$ | Parameters | Type[†] |
|---|---|---|---|
| 1 / Omori [1894] | $k(t+h)^{-1}$ | $k, h$ | Power (OL) |
| 1 / Omori [1894] | $k(t+h)^{-1}+k'(t+h)^{-2}$ | $k, h, k', ...$ | Power series |
| 2[*] / Hirano [1924] | $b(t+a)^{-c}$ | $b, a, c$ | Power (MOL) |
| 3[*] / Jeffreys [1938] | $K(t-\beta)^{(k-1)}$ | $K, \beta, k$ | Power (MOL) |
| 4 / Utsu [1961] | $K(t+c)^{-p}$ | $K, c, p$ | Power (MOL) |
| 5 / Mogi [1962] | $\{E_0 t^{-h}, n_0 e^{-pt}\}$ | $E_0, h, n_0, p$ | Piecewise power / exp. |
| 6 / Burridge and Knopoff [1967] | $ce^{-\alpha t}$ | $c, \alpha$ | Exp. |
| 7 / Souriau et al. [1982] | $\alpha \beta t^{(\beta-1)} e^{(-\alpha t^\beta)}$ | $\alpha, \beta$ | Stretched |
| 8 / Ogata [1983] | $K_1(t+c_1)^{-p_1}+H(t-T_2)K_2(t-T_2+c_2)^{-p_2}+H(t-T_3)K_3(t-T_3+c_3)^{-p_3}+...$ | $K_1, c_1, p_1, K_2, c_2, p_2, K_3, c_3, p_3, ...$ | Epidemic (MOL) |
| 9[*] / Otsuka [1985] | $Ke^{-\alpha t}(t+c)^{-p}$ | $K, \alpha, c, p$ | Power (MOL) with cutoff |
| 10 / Yamashita and Knopoff [1987] | $At^{-p}$ | $A, p$ | Power |
| 10 / Yamashita and Knopoff [1987] | $At^{-1-(\gamma-1)/q}$ | $A, \gamma, q$ | Power |
| 11 / Ogata [1988] | $\sum_i K(t-t_i+c)^{-p} e^{\beta(m_i-M_r)}$ | $K, c, p, \beta, M_r$ | Epidemic (ETAS) |
| 12 / Kisslinger [1993] | $qN^*/t \, (t/t_0)^q \, e^{-(t/t_0)^q}$ | $q, N^*, t_0$ | Stretched |
| 13 / Dieterich [1994] | $r/\{e^{-\Delta\tau/A\sigma}+t\dot{\tau}_r/A\sigma\}$ | $r, \Delta\tau, A\sigma, \dot{\tau}_r$ | Power (OL) |
| 14 / Gross and Kisslinger [1994] | $qN^*e^{(d/t_0)^q} \, 1/(t+d) \, \{(t+d)/t_0\}^q \, e^{-\{(t+d)/t_0\}^q}$ | $q, N^*, d, t_0$ | Stretched |
| 15 / Helmstetter and Sornette [2002] | $(b-\alpha)/b \, 10^{\alpha(m-m_0)}/(1-n) \, t^{*(-\theta)}/t^{(1-\theta)} \sum_k (-1)^k (t/t^*)^{k\theta}/\Gamma\{(k+1)\theta\}$ | $b, \alpha, m_0, n, t^*, \theta$ | Gamma |
| 16 / Narteau et al. [2002] | $A\{\Gamma(q,\lambda_b t)-\Gamma(q,\lambda_a t)\}/t^q$ | $A, q, \lambda_b, \lambda_a$ | Gamma |
| 17 / Lolli and Gasperini [2006] | $A\{\Gamma(q)-\gamma(q,\lambda_a t)\}/t^q$ | $A, q, \lambda_a$ | Gamma |
| 18 / Ben-Zion and Lyakhovsky [2006] | $N_0/\{2\phi R(1-\alpha_s)N_0 t+1\}$ | $N_0, \phi, R, \alpha_s$ | Power (OL) |

* Formulas taken from Utsu et al. [1995]; † Power: power law, Epidemic: epidemic-type power law, Exp.: exponential, Stretched: stretched exponential, Gamma: Gamma-related, OL: Omori law, MOL: Modified Omori law, ETAS: Epidemic-Type Aftershock Sequence model.

**Table 2.** Optimized model percentage of success.

|  | Southern California | Northern California | Taiwan |
|---|---|---|---|
| Nearest-neighbor method [Zaliapin et al., 2008] | | | |
| Number of sequences ($m_{min}$) | 25 (2.6) | 16 (2.4) | 26 (2.5) |
| Power law ($t_{min}$) | 4% (1.00) | 6% (0.82) | 4% (0.63) |
| Exponential ($t_{min}$) | 0% (1.00) | 6% (1.00) | 4% (0.82) |
| Stretched exp. ($t_{min}$) | **100%** (0.04) | **94%** (0.04) | **96%** (0.02) |
| Second-order moment method [Reasenberg, 1985] | | | |
| Number of sequences ($m_{min}$) | 20 (2.6) | 13 (2.4) | 12 (2.8) |
| Power law ($t_{min}$) | 0% (0.40) | 0% (0.40) | 0% (0.32) |
| Exponential ($t_{min}$) | 20% (1.00) | 38% (1.00) | 42% (0.82) |
| Stretched exp. ($t_{min}$) | **90%** (0.08) | **85%** (0.03) | **100%** (0.03) |
| Window method [Gardner and Knopoff, 1974] | | | |
| Number of sequences ($m_{min}$) | 36 (2.4) | 23 (2.3) | 74 (2.6) |
| Power law ($t_{min}$) | 0% (1.00) | 0% (0.63) | 0% (1.00) |
| Exponential ($t_{min}$) | 42% (0.63) | 39% (0.25) | **66%** (0.40) |
| Stretched exp. ($t_{min}$) | **61%** (0.10) | **74%** (0.16) | 34% (0.16) |